\documentclass[apj]{emulateapj}

\usepackage{graphicx}
\usepackage{apjfonts}

\newcommand{\hess}{H.E.S.S.}
\newcommand{\magic}{MAGIC}
\newcommand{\ver}{VERITAS}
\newcommand{\mm}{M87}

\begin{document}

\title{Interpretation of the flares of M87 at TeV energies in the cloud-jet interaction scenario}


\author{Maxim V.~Barkov$^{1,2}$, 
Valent\'i Bosch-Ramon$^{3}$
and Felix A.~Aharonian$^{1,3}$}

\affil{$^1$Max-Planck-Institut f\"ur Kernphysik, Saupfercheckweg 1, 69117 Heidelberg, Germany\footnote{bmv@mpi-hd.mpg.de }\\
   $^2$Space Research Institute, 84/32 Profsoyuznaya Street, Moscow, 117997, Russia \\
   $^3$Dublin Institute for Advanced Studies, 31 Fitzwilliam Place, Dublin 2, Ireland}


\begin{abstract}
{Active galactic nuclei with misaligned jets have been recently established as a class of high-energy gamma-ray sources.
M87, a nearby representative of this class, shows fast TeV variability on timescales less than one day. We present
calculations performed in the framework of the scenario in which gamma-ray flares in non-blazar active galactic nuclei
are produced by a red giant or a gas cloud interacting with the jet. We show that both the light curve and energy
spectrum of the spectacular April 2010 flare can be reproduced by this model, assuming that a relatively massive cloud of
$\sim 10^{29}$~g penetrates into the jet at few tens of Schwarzschild radii from the super-massive black hole.}

\end{abstract}


\maketitle  

\section{Introduction}
\label{intro}

The nearby radio galaxy \mm\, is a unique source for studies of the physics of relativistic plasma outflows and the
conditions in the surroundings of super-massive black holes (SMBH). Because of its proximity ($16.7 \pm 0.2$\,Mpc;
\citealt{mbc07}) and the very massive black hole at its center, with mass $M_{\rm BH} \simeq (3 - 6) \times
10^{9}$\,M$_\odot$ \citep{mma97,gt09}, high resolution very long baseline interferometry (VLBI) at radio wavelengths enables
one to directly probe structures with sizes down to $<100$ Schwarzschild radii ($R_{\rm Sch}$). From the detection
of
super-luminal features in the jet at optical and radio wavelengths, it has been possible to constrain the jet orientation
angle towards the line of sight on sub-kpc scales to $\theta \lesssim 20^\circ$ \citep{bsm99,chs07}. The source 
is considered to be a misaligned jet active galactic nuclei (AGN).

\mm\, shows very high-energy (VHE) recurrent activity with variability
timescales of a few days or less \citep{ah06,mag,vvhm09,ver10}. In April 2010, a bright VHE gamma-ray flare was
simultaneously detected by three ground based Cherenkov telescopes: \hess, \magic\, and \ver\, \citep{hess11}.
The detection of the VHE flare triggered further observations in X-rays ({\it Chandra}), and radio (43\,GHz; VLBA). The
excellent sampling of the VHE gamma-ray lightcurve allows a precise temporal characterization of the flare, which is well
described by a two-sided exponential function with significantly different flux rise and decay times of
$\tau_{\mbox{d}}^{\mbox{rise}} = 2.9$\,days and  $\tau_{\mbox{d}}^{\mbox{decay}} = 0.9 $\,days, respectively. The peak flux
was $\Phi_{>0.35\,{\rm TeV}} \simeq 2 \times 10^{-11}$\,ph\,cm$^{-2}$\,s$^{-1}$, which given the unbeamed nature of the
radiation allows the derivation of a safe estimate of the luminosity, $L_{\gamma}\sim 10^{42}$~erg~s$^{-1}$. X-ray {\it
Chandra} observations taken approximately $2-3$ days after the peak of the VHE gamma-ray emission revealed an enhanced flux
from the core by a factor of $\sim 2$ ($L_{\rm X} \sim 10^{41}$~erg~s$^{-1}$), with a variability timescale of $<2$ days
\citep{hess11,zry11}. \ver\, obtained VHE spectra consistent with a power-law for three flare phases: rising flux, peak flux,
and falling flux \citep{veritas11}. At the peak of the flare, the photon index was $\approx 2.2$, and there is indication at
a few $\sigma$ level that the spectrum is somewhat softer in the rising and falling phases. {\it Fermi} has detected the
source above 100~MeV with a luminosity $\approx 5\times 10^{41}$~erg~s$^{-1}$ \citep{fer092}. However, given the M87 flux and
the {\it Fermi} sensitivity, this instrument could not probe day-scale variability in this source.

Several theoretical scenarios have been proposed to explain the TeV flares in \mm. \cite{gpk05} and \cite{lbs08}
showed that one-zone (homogeneous) leptonic synchrotron self-Compton (SSC)
models are unlikely to explain the observed TeV spectrum of \mm. There are also
leptonic jet models with a more complex emitter. In a multi-blob scenario \citep{lbs08}, a low
magnetic
field in the emitting region is required, which may be at odds with the fact that these
regions of the jet are likely strongly magnetized \citep[e.g.,][]{kbvk07,bk08}. This problem may be overcome if the
acceleration and/or the emission processes take place in a weakly magnetized cloud rather than in the jet\footnote{Magnetic
diffusion longer than the flare timescale would prevent the jet magnetic field from penetrating into the cloud.}. There is also
the spine-sheath model \citep{tg08}, which predicts very strong gamma-ray absorption if it tries to explain the VHE hard spectra and
the fast variability. Another possibility is the jet-in-jet model of \cite{gub10}, which can reproduce the spectrum of the
2010 flare, but does not provide at this stage a quantitative prediction of the lightcurve of the flare. In the work by
\cite{cylw11}, the authors may explain the VHE flares with an external inverse Compton model, but they need to assume a very
wide jet to be able to invoke Doppler boosting. A SMBH magnetospheric origin for the TeV emission in M87 has been also
proposed \citep{na07,ra08,vl10,lr11}. Magnetospheric models may explain a hard spectrum at VHE, but there is at present
no detailed
quantitative prediction for the VHE lightcurve. Synchrotron-proton emission may also operate in M87
\citep[][]{rpd04}, but to explain the extension of the gamma-ray spectrum beyond $\sim 1$~TeV, strong Doppler boosting is
needed, which is not the case in this source. Finally, in the paper by \cite{bab10} the TeV flares were explained by the
interaction between the \mm\, jet and a dense gas cloud formed out of the disrupted atmosphere of a red giant (RG). The
emission is produced by proton-proton ($pp$) interactions between protons accelerated at the jet-cloud interface. This model
predicts a hard gamma-ray spectrum from GeV to TeV energies and fast variability on timescales of days. Unlike in models in
which the VHE emission takes place in the dilute jet matter, $pp$ interactions in the dense cloud can be energetically very
efficient, and the short dynamical timescale provides with the fast variability required to explain the flares in \mm.

In this work, we develop further the scenario presented in \cite{bab10}. We demonstrate that this model can naturally explain
both the very short variability and the gamma-ray spectrum as detected. In Sect.~\ref{model}, we describe the model, the
radiation features of which are explored in Sects.~\ref{nt} and \ref{sed}, and discussed in Sect.~\ref{disc}.

\section{The model}
\label{model}

\begin{figure}
\includegraphics[width=84mm,angle=-0]{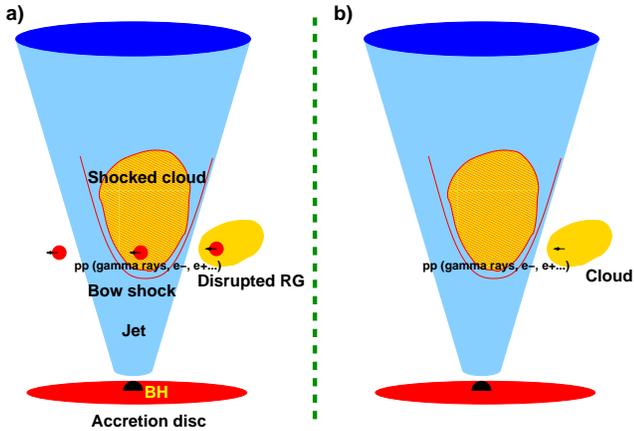}
\caption{Sketch of the considered scenario; a) penetration of an RG, with the external layers detached 
due to gravitational disruption, into the jet; b) penetration of a massive clump of matter into the jet.
This sketch has been adapted, with minor modifications, from \cite{bpb12}.}
\label{sk}
\end{figure}

The model considered here is based on the scenario proposed by \cite{bab10}, in which the envelope of an RG, partially
tidally disrupted by the SMBH gravitational field, is shocked by the jet and torn away from the stellar core. Due to the
jet impact, the RG envelope is blown up by the jet ram pressure, forming a cloud of gas heated and accelerated
downstream \citep[see also][]{bpb12}. In the present work, we will consider a generic gas cloud sufficiently massive
to have a dynamical impact on the jet, at least temporally, so it can tap a substantial fraction of the jet luminosity.
The interaction of the jet with such a cloud can convert a part of the jet magnetic and kinetic energy into internal
energy, and a significant fraction of it could go to accelerate protons and electrons. Given the large magnetic fields
expected in the jet base \citep[e.g.,][]{kbvk07,bk08}, electrons are unlikely to reach TeV emitting energies unless the
accelerator is screened from the jet magnetic field, whereas protons do not suffer from this limitation. A significant
part of the accelerated protons can reach the cloud, in which case optically-thick $pp$ collisions will lead to
significant gamma-ray production in the early stages of the cloud expansion. A sketch of the scenario considered here is
shown in Fig.~\ref{sk}.

The atmosphere of an RG { provides} a good target for the jet to interact close to the SMBH, which is not the case for
the stellar atmospheres of main sequence stars. The reason is that RGs have external layers that are much less
gravitationally bounded to the stellar core. In the vicinity of a SMBH, the external layers of an RG will suffer
significant tidal disruption \citep[see][]{knp93b,knp93a,dfkn97,alp00,icn03,lkp09}, and a mass as high as $\gtrsim
10^{30}$~g can be left almost gravitationally unbound. Therefore, if an RG penetrates into the innermost region of the
jet, it can suffer the loss of its external layers due to jet ablation. Without gravitational disruption, the mass loss
will be significantly reduced except for very powerful jets \citep[see][]{babkk10}. Winds from stars could be effective
clumps for the jet as well, but at larger distances from the SMBH, implying thus longer dynamical timescales.

We note that, in addition to disrupted RG envelopes, other types of matter clumps could also be considered, like dark
and ionized clouds in weak and powerful active galactic nuclei, respectively \citep[see, e.g.,][and references
therein]{abr10}. One should mention in this context that a cloud of $\sim 10^{28}$~g was detected in the center of our
Galaxy \citep{ggf12}.

\subsection{The cloud dynamics}
\label{dyn}

One of the key  parameters of the model is the power of the jet, which in M87 is $L_{\rm j}\approx (1-5)\times
10^{44}$~erg~s$^{-1}$ \citep{oek00}. In this work, we fix this value to $L_{\rm j}\approx 5\times 10^{44}$~erg~s$^{-1}$.
From $L_{\rm j}$ and the jet radius, $R_{\rm j}=\theta\,z_{\rm j}$, we can derive the jet energy flux at the interaction
height $z_{\rm j}$:
\begin{equation}
F_{\rm j}=\frac{L_{\rm j}}{\pi z_{\rm j}^2\theta^2} \approx 3\times 10^{13}\;
L_{\rm j,44} z_{16}^{-2} \theta_{-1}^{-2} \mbox{ erg~cm}^{-2} \mbox{ s}^{-1}.
\label{lir}
\end{equation}
where $\theta_{-1}=\theta/0.1$ is the jet semi-opening angle in radians, and $z_{16}=(z_{\rm j}/10^{16}\,{\rm cm})$ 
is the distance from the SMBH at which the cloud crosses the jet. 

In the RG case, there are two tidal disruption regimes. Under strong tidal interaction, the radius of the RG, $R_{\rm
RG}$, is larger than the tidal disruption radius of the star, $R_{\rm *T}$ (see Eq.~2 in \citealt{bab10}). In that case,
the RG envelope becomes elongated along the direction of motion of the star \citep{knp93b}. Under weak tidal
interaction, when $R_{\rm RG}\sim R_{\rm *T}$, the envelope is still roughly spherical \citep{knp93a}. In both
situations, the outer layers of the star will be swept away by the jet, forming a cloud that will be quickly heated up
and expand. 

We study the time evolution of the cloud adopting a very simplified hydrodynamical model for the cloud expansion. The
heating of the cloud is caused by the propagation of shock waves, which are formed by the pressure exerted by the jet
from below. { Therefore, the cloud pressure is taken similar to the jet ram pressure (regardless it is of kinetic or
magnetic nature) $ p_{\rm j}={F_{\rm j}}/{c}\approx p_{\rm c}\approx(\hat\gamma-1)e_{\rm c}\,,$
where $c$ is the speed of light, and $\hat\gamma$ the adiabatic index, fixed here to $4/3$ because the cloud is, at
least initially, radiation dominated \citep[see][]{bab10}. The cloud expands sidewards at its sound speed ($c_{\rm s}$),
since the external pressure is mostly exerted on the cloud bottom, i.e. from the jet upstream direction
\citep[e.g.][]{phf10,bpb12}. As expansion proceeds, the cloud pressure becomes smaller}; when it is lower than the
jet ram pressure, new shocks develop in the contact discontinuity, heating further the cloud and thus accelerating its
expansion. We are not concerned here with the latest expansion phase at which the jet lateral pressure may confine the
cloud. We do not take into account either the energy transfer due to cosmic rays entering in the cloud, which would
enhance the late expansion rate.

Numerical calculations \citep[e.g.][]{gmr00,nmk06,phf10,bpb12} show that the cloud is destroyed on few cloud-crossing
times, $r_{\rm c0}/c_{\rm s0}$, where $r_{\rm c0}$ and $c_{\rm s0}$ are correspondingly the initial radius and  sound
speed of the cloud after the first shock wave has crossed it. These simulations also show that the radius of the volume
containing fragments of the destroyed cloud can grow up to an order of magnitude compared to $r_{\rm c0}$ \citep[see
Fig.~15 in][]{bpb12}. The fragmented cloud, with a velocity still different from that of the jet, continues to be
suitable for shock formation and particle acceleration. The assumption of a spherical cloud, as adopted here, is  a
simplification, but it allows an analytical treatment of such a complicated system. In what follows, we present,
depending on the tidal strength, possible analytical dynamical models of the cloud to be used to compute the jet-cloud
interaction radiation.

\subsubsection{Weak tidal interaction }
  
The solution of the system of equations for the cloud radius evolution with time, $t$,
in the weak tidal interaction case, can be written as follows \citep[see more details in ][]{bab10}:
\begin{equation}
r_{\rm c}(t)=\frac{r_{\rm c0}}{(1-t/t_{\rm ce})^2 }
\label{rcl_j}
\end{equation} 
where $r_{\rm c0}$, assumed to be similar to $R_{\rm *T}$, 
is the initial cloud radius, and $t_{\rm ce}$ the cloud characteristic 
expansion time:
\begin{equation}
t_{\rm ce} = \left(\frac{3 c M_{\rm c}}{\pi\hat\gamma F_{\rm j}r_{\rm c0}}\right)^{1/2}
\approx 5 \left(M_{\rm c28}/F_{j,14}r_{\rm c0,13}\right)^{1/2}\,{\rm days}\,,
\label{Acl}
\end{equation}
where $M_{\rm c28}=M_{\rm c}/10^{28}\,{\rm g}$. Since the velocity along the jet direction is non-relativistic during
the time of interest, we neglect this motion component.

\subsubsection{Strong tidal interaction}

In the case of strong tidal interaction the RG atmosphere is stretched in the direction of motion of the star, and the
expansion will be now cylindrical; this would also apply to other types of clumps (see below).
In such a case, the initial cloud cylindrical radius $r_{\rm c0}$ 
can be significantly smaller than the length of the disrupted atmosphere, $l_{\rm c}$ \citep{alp00}. { The solution 
describing this case can be written as $r_{\rm c}(t)=r_{\rm c0} e^{t/t_{\rm ce}}.$}
As in the weak case, $r_{\rm c0}$ and $t_{\rm ce}$ are the initial radius and the expansion time of the cloud, where:
\begin{equation}
t_{\rm ce}=\left(\frac{cM_{\rm c}}{\pi \hat\gamma F_{\rm j} l_{\rm c}} \right)^{1/2}
=1 M_{\rm c,28}^{1/2} F_{\rm j,14}^{-1/2} l_{\rm c,14}^{-1/2}\,\mbox{day}\,,
\label{tde}
\end{equation} 
with $l_{\rm c,14}=(l_{\rm c}/10^{14}\,{\rm cm})$ and $F_{\rm j,14}=(F_{\rm j}/10^{14}\,
{\rm erg~cm}^{-2}{\rm s}^{-1})$.

\begin{figure}
\includegraphics[width=70mm,angle=-90]{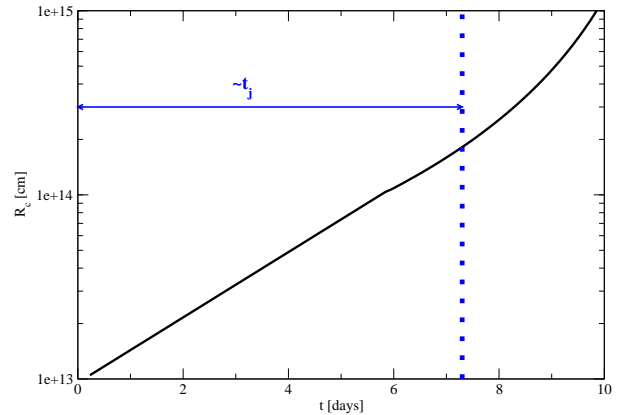}
\caption{Cloud radius evolution for the elongated and spherical cases combined. 
The parameter values are: $F_{\rm j}=1.6\times 10^{14}$~erg~cm$^{-2}$~s$^{-1}$, $r_{\rm c}=10^{13}$~cm, 
$l_{\rm c}=2\times 10^{14}$~cm, and $M_{\rm c}=2\times 10^{29}$~g. The vertical dotted line ($t_{\rm j}$) indicates the
moment when the cloud leaves the jet.}
\label{rct}
\end{figure}

As mentioned above, in case the clump were not of RG origin, the initial cloud shape would be more uncertain given the lack of a
stellar core. It seems reasonable however to assume that the cloud will be elongated by tidal forces and fast motion in the intense 
SMBH gravitational field. Therefore, in the next section we adopt the dynamics of a strong tidal interaction, and when $r_{\rm
c}=l_{\rm c}$ we switch to the spherical solution. In Fig.~\ref{rct}, the evolution of the radius is shown for the following 
characteristic parameter values: $F_{\rm j}=1.6\times 10^{14}$~erg~cm$^{-2}$~s$^{-1}$, $r_{\rm c}=10^{13}$~cm,  $l_{\rm c}=2\times
10^{14}$~cm, and $M_{\rm c}=2\times 10^{29}$~g. The value of $F_{\rm j}$ has been derived assuming $L_{\rm j}=5\times
10^{44}$~erg~s$^{-1}$, $\theta=2^\circ$ and $z_{\rm j}=3\times 10^{16}$~cm, about 20~$R_{\rm Sch}$ for $M_{\rm BH}=6\times
10^9\,M_\odot$. The time at which the cloud would leave the jet is also indicated in the figure.

\section{Radiation processes}\label{nt}

\subsection{VHE gamma rays from $pp$ collisions}

As noted in Sect.~\ref{intro}, it seems likely that the jet is
still magnetically dominated at $z\sim z_{\rm j}$. One can estimate the magnetic field in the jet at the level of
$\sim 100$~G at a distance from the black hole of $\sim 10^{16}$~cm.  Such a high
magnetic field, if not dissipated, will prevent the formation of a hydrodynamical shock. However, the $B$-field could
have an alternating polarity \citep[e.g.][]{lyu10}, and compression against the clump may lead to very effective
magnetic reconnection, a potential mechanism to accelerate particles (e.g. \citealt{zen01}; see also \citealt{bos12} for the context of obstacle-jet interactions). Magnetic
reconnection can also lead to the eventual development of a hydrodynamic shock and strong turbulence
\citep[e.g.][]{sir11a}, both giving rise to Fermi~I and II type acceleration
processes and to magnetic field suppression.
Given the complexities of the processes at play,
we postulate here that particles are accelerated in the jet-cloud interaction region without specifying the acceleration
mechanism. 
%
The proton energies can easily reach $\sim 1$~PeV. Even
for a diffusion regime faster than Bohm diffusion, protons could still reach high enough energies to explain
the observations. 

As shown in \cite{bab10}, photo-meson production and proton synchrotron will not be efficient, but the cloud
density can be high, making $pp$ interactions the best
channel for gamma-ray production in cloud-jet interactions in \mm. The characteristic cooling time for $pp$ collisions is:
\begin{equation}
t_{pp}\approx \frac{10^{15}}{c_{\rm f}n_{\rm c}}=10^5\,c_{\rm f}^{-1} M_{\rm c,28}^{-1}\,r_{\rm c,14}^{3}\,{\rm s}\,,
\label{pp}
\end{equation}
where $n_{\rm c}$ is the spherical cloud density, $r_{\rm c,14}=(r_{\rm c}/10^{14}\,{\rm cm})$, and $c_{\rm f}\approx (\hat\gamma+1)/(\hat\gamma-1)$ is a constant that
takes into account the cloud compression by the jet shock (neglecting post-shock radiative cooling).

In $pp$ interactions, the fraction of the proton energy transferred per collision to the leading $\pi^0$-meson, which 
yields two photons, 
is $\approx 0.17$ \citep{kab06}; this value in the optically thick case is larger by a factor of 2.
Therefore, one can characterize the proton-to-gamma-ray energy transfer by
\begin{equation}
\chi\equiv E_{\gamma}/E_p = 0.17\,[2-\exp(-t/t_{pp})]\,,
\label{enef}
\end{equation}
where $t$ can be fixed to the characteristic time of variability ($t_{\rm v}$), since $t_{\rm pp}$ grows much faster
with time.

Two phases of the cloud expansion can be distinguished: the radiatively efficient regime, i.e. with  $\chi\approx 0.34$
or $t_{pp}<t_{\rm v}$, and the radiatively inefficient regime, with $\chi=0.17$ or $t_{pp}>t_{\rm v}$. 
Thus, from the simplifications
above, the gamma-ray luminosity in the $pp$ optically-thick case can be written as:
\begin{equation}
L_{\gamma}\approx 0.34\,\eta\,\pi\,r_{\rm c}^2\,F_{\rm j}\,,
\label{egc}
\end{equation} 
where it has been assumed that the injected proton luminosity, $L_p$, grows like the jet-cloud interaction 
surface, $\propto
r_{\rm c}^2$. The parameter $\eta$ is the efficiency of energy conversion from the jet to the accelerated particles. 
In the optically-thin regime, only a fraction $\sim t_{\rm v}/t_{pp}$ of $L_p$ is lost through $pp$ collisions,
and  $L_{\gamma}\propto r_{\rm c}^{-1}$. The gamma-ray luminosity during the cloud-jet interaction is:
\begin{equation}
L_{\gamma}\approx \pi\eta\chi\,r_{\rm c}^2\,F_{\rm j}\,\left(1-e^{-t_{\rm v}/t_{pp}}\right)\,. 
\label{egcm}
\end{equation} 
Given the fast expansion of the cloud, either in the spherical or the elongated case, one can expect a sharp spike in the
lightcurve. Equation~(\ref{egcm}) gives a smoother lightcurve peak than Eq.~(32) in \cite{bab10} because of new
exponential shape and the decay phase is now constraint by $t_{\rm v}$, a modification supported by the numerical
calculations of Section~\ref{sed}.

\begin{figure}
\includegraphics[width=84mm,angle=-0]{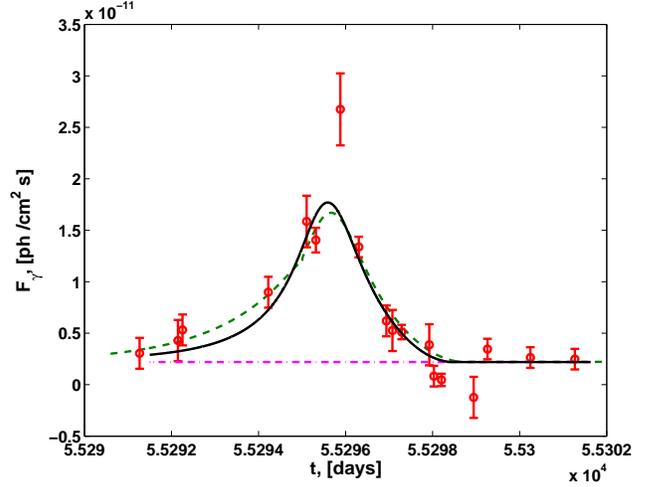}
\caption{Computed lightcurve of the Gamma-ray $pp$ photons for the weak tidal disruption case (solid line). The
time convention is MJD. The
parameter values are chosen for \mm: $F_{\rm j}=1.6\times 10^{14}$~erg~cm$^{-2}$~s$^{-1}$, $r_{\rm c0}=3\times 10^{13}$~cm,
$c_{\rm f}=7$, $\eta=0.3$, $t_{\rm v}=7\times 10^4 \mbox{ s}$, and $M_{\rm c}\approx 10^{29}$~g.
A small pedestal flux of $\sim 2\times 10^{-12}$~ph~cm$^{-2}$~s$^{-1}$ has been added (dash-dotted line).
The lightcurve for the strong (elongated) tidal disruption case (dashed line)
is also shown, for which $t_{\rm v}=9\times 10^4 \mbox{s}$, $r_{\rm c0}=10^{13}$~cm, and $l_{\rm c}=2\times
10^{14}$~cm.}
\label{degdtsc}
\end{figure}

Analytical lightcurves similar to that derived in \cite{bab10} are presented in Fig.~\ref{degdtsc} for
the
weak and the strong tidal interaction cases.  The adopted parameter values for the
weak  tidal interaction case are: $c_{\rm
f}=7$, $t_{\rm v}=7\times 10^4 \mbox{ s}$, $M_{\rm c}\approx 10^{29}$~g, and $F_{\rm j}=1.6\times
10^{14}$~erg~cm$^{-2}$~s$^{-1}$. As seen from the figure, the VHE peak is reached at $t_{\rm peak}\approx 4\times 10^5$~s,
with a width of $\sim 1-2$~days,  adopting $\eta=0.3$. To obtain a short rise phase we use a relatively large cloud initial
radius: $r_{\rm c0}=3\times10^{13}$~cm. 
In the strong tidal interaction case one can set a smaller initial radius of the RG, since the clump is already elongated due
to tidal forces before penetrating into the jet. The adopted parameter values are the same as in weak interaction case but $t_{\rm v}$, fixed now to
$9\times
10^4 \mbox{ s}$, and $r_{\rm c0}=10^{13}$~cm and $l_{\rm c}=2\times 10^{14}$~cm. 
As seen in the figure, the VHE peak is reached at $t_{\rm peak}\approx 5\times 10^5$~s, with a width of
$\sim 1-2$~days.


\subsection{Non-thermal electrons and positrons}

Secondary electrons and positrons ($e^\pm$) are injected into the cloud through the decay of charged secondary $\pi$-mesons produced in
$pp$ collisions. The energy rate of the injected $e^\pm$ pairs is $\approx
L_\gamma/2$ \citep{kab06}. The region is quite compact, and depending on the cloud magnetic
field most of the secondary emission can be synchrotron or SSC, with a minor contribution from 
relativistic
Bremsstrahlung. The initial magnetic field in the cloud can be relatively small; e.g. in the case of a (non-disrupted) RG
atmosphere, the magnetic field is of several gauss \citep[e.g.][]{kon08,kon10}. The quick cloud expansion can decrease the 
$B$-strength
rapidly, although the continuous pumping of jet energy, plus complex dynamo effects in the cloud, 
may prevent the $B$-field from
decreasing, and may even enhance it. Assuming that X-rays right after the peak come from the cloud, an estimate of
the cloud magnetic field can be done assuming that synchrotron losses dominate over SSC. 
This yields a lower limit for the cloud magnetic field ($B_{\rm
c}$), since its energy density, $B_{\rm c}^2/8\pi$, is to be larger than the synchrotron one, $\varpi_{\rm syn}\lesssim
L_{\rm \gamma X}/4\pi r_{\rm c}^2c$, where $L_{\rm \gamma X}$ is the gamma-ray luminosity at the X-ray observation. This yields a (loose) constraint on $B_{\rm c}$:
\begin{equation}
B_{c}\sim \left({8\pi\varpi_{\rm syn}}\right)^{1/2}\sim 30\,L_{\gamma X,41}^{1/2}r_{\rm c,14}^{-1}\,{\rm G}\,, 
\label{bx} 
\end{equation}
where $L_{\rm \gamma X,41}=(L_{\rm \gamma X}/3\times 10^{41}\,{\rm erg~s}^{-1})$. Note that even for lower $B_{\rm
c}$-values, gamma-rays could be absorbed through photon-photon ($\gamma\gamma$) collisions in the synchrotron field.
Given the different possible $B_{\rm c}$-values and radiation outcomes, 
we will treat $B_{\rm c}$ after the cloud expansion as a free
parameter confined to a value range $\sim 0.01-100$~G.

Primary $e^\pm$ pairs carried by the jet may be also accelerated together with protons at the jet-cloud contact
discontinuity. Since their losses are expected to be synchrotron dominated given the high $B_{\rm j}$-value, their emission
should not overcome the X-ray fluxes detected few days after the VHE peak. The primary synchrotron radiation should not
absorb the VHE photons either. All this sets an upper limit on the primary
$e^\pm$ injection rate of $\sim 0.01\,L_p$ around the VHE maximum. Beside this restriction, however, the primary leptonic
population is rather unconstrained, whereas secondary pairs are almost completely constrained by the relativistic proton
spectrum and estimated cloud conditions, with only the $B_{\rm c}$-value remaining free. Below we examine under which
conditions these secondaries do not lead to strong gamma-ray absorption, and explore whether they may still explain the
enhanced X-ray emission detected few days after the VHE maximum.

\section{Modeling the high-energy emission}\label{sed}

The emitting proton population has been modelled adopting a spatially homogeneous (one-zone) emitter, in which relativistic
protons are injected with power-law energy distributions of two types: $Q_p(E)\propto E^{-2}$ plus an exponential cutoff at
$E_{\rm c}= 200$~TeV and a low-energy cutoff at $E_{\rm LE}=1$~TeV; and $Q_p(E)\propto E^{-1.5}$ plus an exponential cutoff
at $E_{\rm c}=50$~TeV. The first case would correspond to a fairly standard phenomenological assumption plus a low-energy
cutoff, and the second one may be associated to non-linear shock acceleration with a large compression ratio.

The injection luminosity of protons has been taken $L_p=\eta\,\pi\,r_{\rm c}(t)^2\,F_{\rm j}$, and thus the model is inhomogeneous in
time. To derive the proton energy distribution for different $t$-values ($N(E,t)$), the time-dependent proton injection and cloud
conditions have been modelled as follows. Protons injected at a certain time $t_i$, $Q_p^i(E)$, are evolved for a $\delta t_i$ under
the corresponding cloud conditions. The evolved population is added to the accumulated population from $t_1$ to $t_{i-1}$ evolved also
for a $\delta t_i$:  $N_{\rm prev}^{\rm i-1}(E)\rightarrow N_{\rm prev}^{\rm i}(E)$. This numerical technique provides a correct
solution at any time of interest provided that $\delta t_i\ll t_{\rm dyn}^i$, where $t_{\rm dyn}^i$ is the dominant evolution
timescale of protons at $t_i$. At the relevant time interval, the proton evolution is dominated by $pp$ interactions over other
radiation cooling mechanisms. Adiabatic losses, given the roughly constant cloud pressure during the relevant expansion phase, do not
play a significant role.

\begin{figure}
\includegraphics[width=70mm,angle=-90]{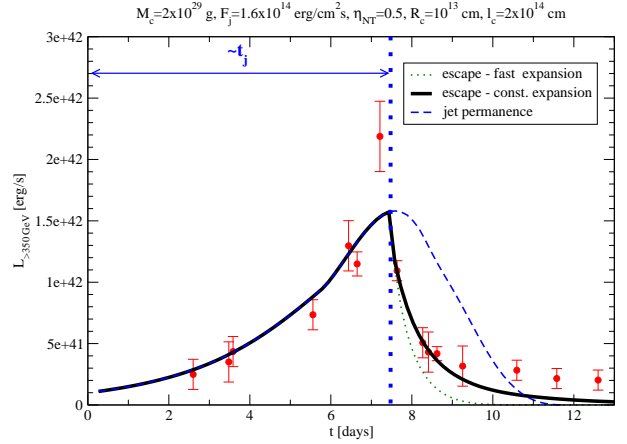}
\caption{Computed VHE-luminosity lightcurve of the flare assuming proton injection 
during 12~days (thin dashed line), and only during 7.3~days (dotted -increasing radius growth speed- and thick
solid line -constant radius growth speed-), as expected if the cloud leaves the jet. 
The parameter values are the same as in Fig.~\ref{rct} plus $\eta=0.5$ and $Q_p(E)\propto E^{-2}$. 
The VHE data points are from \cite{hess11}.}
\label{lcnum}
\end{figure}

In Fig.~\ref{lcnum} we show the VHE lightcurve computed using $N(E,t)$, derived adopting the $Q_p(E)\propto E^{-2}$ case, and a
proton-to-gamma-ray energy fraction per collision of 0.17. The remaining parameter values are the same as in Fig.~\ref{rct} plus a
non-thermal efficiency $\eta=0.5$. The slightly higher $\eta$-value than in Fig.~\ref{degdtsc} comes from the calculation method,
since now the proton emissivity is calculated numerically, whereas in Fig.~\ref{degdtsc} the calculation was analytical. As seen in
Fig.~\ref{lcnum}, the computed VHE peak can be slightly broader than the observed one, although a stronger decay can be naturally
explained by the cloud escaping the jet and shutting off the proton injection. The Keplerian speed at $3\times 10^{16}$~cm ($M_{\rm
BH}=(3-6)\times 10^9\,M_\odot$) is $\approx 4-5\times 10^{9}$~cm~s$^{-1}$, and a cloud with the adopted mass will leave the jet after
several days for $R_{\rm j}=10^{15}$~cm (lighter clouds would be dragged downstream the jet, e.g. \citealt{babkk10,bpb12}). To model
the effect of the cloud escape, two possibilities have been considered together with cutting off the proton injection: either the
cloud expansion continues at the same (growing) rate, or it is kept constant after the escape. 

\begin{figure}
\includegraphics[width=70mm,angle=-90]{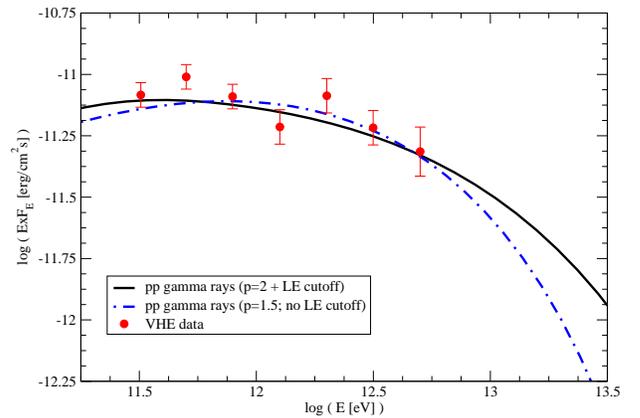}
\caption{The computed SED for $pp$ gamma rays at the VHE maximum for two different injection spectra: $Q_p(E)\propto E^{-2}$ 
($E_{\rm LE}=1$~TeV, $E_{\rm c}=200$~TeV; solid line); 
and $Q_p(E)\propto E^{-1.5}$ (no $E_{\rm LE}$, $E_{\rm c}=50$~TeV; dot-dashed line). 
The remaining parameter values are the same as in Fig.~\ref{lcnum}. The VHE data points are from \cite{veritas11}.}
\label{sedpp}
\end{figure}

The VHE spectral energy distributions (SED) obtained adopting $Q_p(E)\propto E^{-2}$ and $E^{-1.5}$ are presented in Fig.~\ref{sedpp}.
Despite the hard proton spectrum, the lower energy exponential cutoff of the case with $Q_p(E)\propto E^{-1.5}$ renders a very similar
$pp$ SED at VHE.

Some remarks should be done regarding the energetics and the physical consistency of the discussed dynamical-radiation model. For
either the $Q_p(E)\propto E^{-2}$ or the $Q_p(E)\propto E^{-1.5}$ case, the required $L_{\rm p}$-value to explain the VHE event in the
context of the adopted dynamical model amounts $\sim 10^{43}$~erg~s$^{-1}$.  This means that the average efficiency of
proton-to-photon energy transfer is $\sim 15$\%. This also implies a rather tight energy budget, $\eta=0.5$.  For the case with
$Q_p(E)\propto E^{-2}$ with a low-energy cutoff at $\sim$ GeV energies, the luminosity budget would roughly double. This strong energy
requirements are however partially caused by the limitations of the analytical dynamical model. For instance, the compression ratio
has been fixed to $c_{\rm f}=7$, but since expansion eventually makes the cloud optically thin, the radiative nature of the cloud
shocks can render the compression ratio much higher than 7. Another effect to consider here is the cloud fragmentation, which can take
place at the later stages of the cloud expansion \citep[see][]{bpb12}. Cloud fragmentation effectively implies a larger cloud cross
section by a factor $\sim N_{\rm f}^{1/3}$, where $N_{\rm f}$ is the number of cloud fragments. Therefore, the real energetic
requirements may be relaxed by a factor of several. In particular, the very high and sharp peak observed by \hess\, could be explained
when accounting for the aforementioned stronger compression and fragmentation. We did not include however these effects in the present
calculations since this would imply additional free parameters. We also note that the back reaction of accelerated particles, which
can heat the cloud, accelerate its expansion, and shorten the VHE emission rise time, has been neglected. To take into account all
these effects, a more accurate treatment is needed that would provide the coupling between acceleration, radiation and
magnetohydrodynamics (MHD). This will be left for future studies.

\subsection{X-rays}

Few days after the maximum in the VHE lightcurve, {\it Chandra} observed the M87 core and found it in a high-flux state, which
decreased by a factor of two in the following days \citep{zry11}. Although the X-ray emission does not necessarily come from
$pp-e^\pm$ pairs (plus a $\gamma\gamma-e^\pm$ contribution; see below), and may have a primary leptonic origin, it is  interesting to
explore whether $pp$ interactions alone can explain the X-ray and VHE observations. Therefore, together with the $pp$ gamma rays, we
have also calculated the broadband emission of the $pp-e^\pm$ pairs. For that, we derived first the $e^\pm$ injection spectrum
$Q_{e^\pm}(E)$ using the formulae in \cite{kab06}, and computed its evolution under the cloud conditions at the VHE maximum. 

The $e^\pm$ dominant cooling channels within the cloud are synchrotron or SSC, depending on $B_{\rm c}$. Other cooling
channels, like relativistic Bremsstrahlung or inverse Compton with IR photons from the M87 nucleus, are less important. Since
the leptonic cooling processes are quite fast at the relevant energies, the cloud radius and $e^\pm$ pair injection can be
considered as roughly constant. Under SSC dominance, however, the emitter still requires an inhomogeneous treatment in time
because the cooling of $e^\pm$ pairs depends on their own synchrotron radiation, making the evolution non-linear. To account
for this, we have applied to the $e^\pm$ pairs the same numerical technique used for protons. Note that to compute the whole
evolution of the secondary $e^\pm$ pairs and their emission, the $e^\pm$ injection should change in time as the $pp$
gamma-ray luminosity does. As it is computationally rather expensive, we have restricted ourselves to the leptonic SED for
the cloud conditions at the VHE maximum, and adopted two $B_{\rm c}$-values to obtain SEDs for the synchrotron and the SSC
dominated cases.

As discussed in Sect.~\ref{nt}, effective secondary synchrotron radiation, high enough to explain {\it Chandra} April 11th 2010 data,
can suppress the VHE emission. This can happen even for a very hard injected $e^\pm$ population, since the photon energy distribution
under synchrotron cooling below X-rays. In addition, since a relatively high $B_{\rm c}$-value is required for synchrotron radiation
to overcome SSC, electromagnetic cascades will not be able to alleviate the strong gamma-ray absorption. The situation is actually
complex, since in fact the $\gamma\gamma$ pairs can dominate over the pairs from $pp$ interactions, yielding even brighter and harder
X-rays. We find that $B_{\rm c}\lesssim 0.1$~G is necessary to avoid strong absorption, and $B_{\rm c}\gtrsim 10$~G could explain the
X-ray fluxes few days after the VHE maximum.

\begin{figure}
\includegraphics[width=90mm,angle=0]{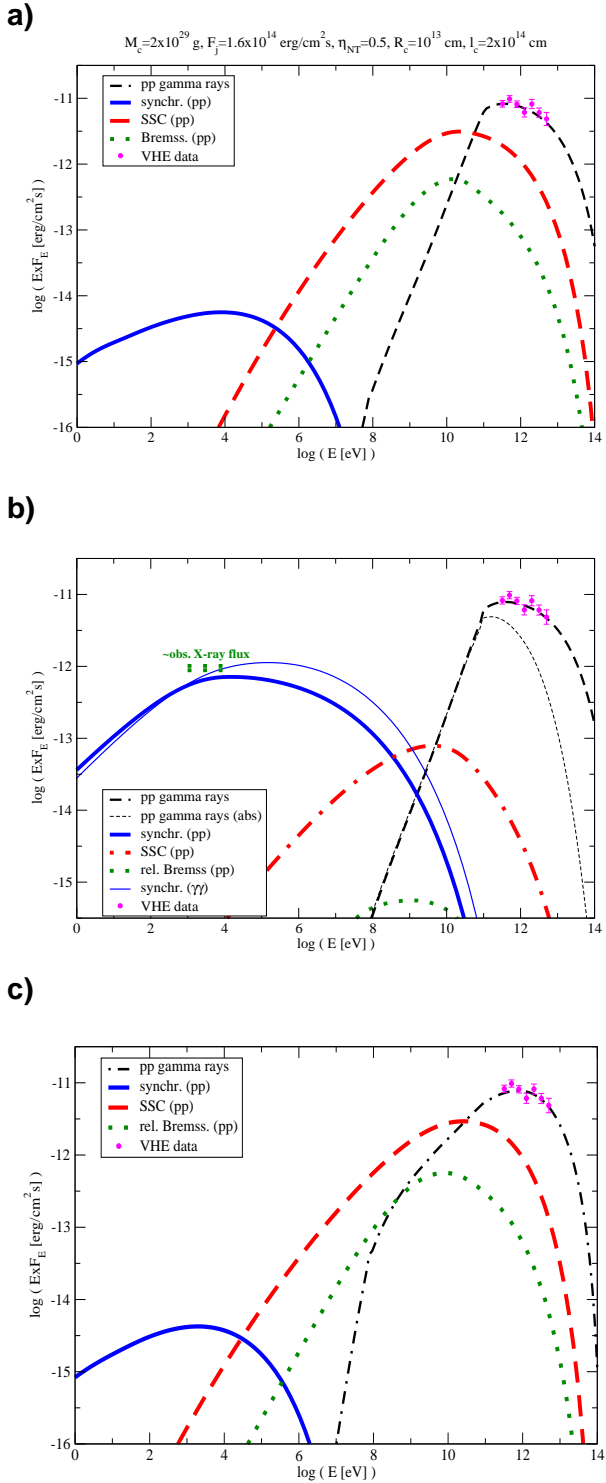}
\caption{{\bf a):} The computed SED for secondary $e^\pm$ synchrotron and SSC, and $pp$ gamma rays, 
adopting $Q_p(E)\propto E^{-2}$, $E_{\rm LE}=1$~TeV, $E_{\rm c}=200$~TeV, 
and $B_{\rm c}=0.03$~G. The other parameters are the same as those adopted to obtain Fig.~\ref{sedpp}.
The VHE data points are from \cite{veritas11}.
{\bf b):} The same as in {\bf a)} but for $B_{\rm c}=30$~G, plus the $\gamma\gamma-e^\pm$ synchrotron emission. 
The approximate level of observed X-ray emission in April 11th 2010 is also shown.
The computed synchrotron emission from the $\gamma\gamma-e^\pm$ pairs are also shown.
{\bf c):} The same as in {\bf a)} but for a proton injection spectrum $\propto E^{-1.5}$, no $E_{\rm LE}$, 
and $E_{\rm c}=50$~TeV.}
\label{sede}
\end{figure}

The computed SED for a low $B_{\rm c}$-value (0.03~G), able to reproduce the observed VHE SED avoiding significant absorption, is
presented in Fig.~\ref{sede} (top) for protons following a $Q_p(E)\propto E^{-2}$. The X-ray high state few days after the VHE peak
can be roughly explained by an increase in $B_{\rm c}$, which may happen via mixing of the strongly magnetized jet material with the
cloud when it fragments, or by dynamo effects in the complex MHD flow. If this were the case, the decay of the VHE lightcurve could
not be explained by jet escape, since high X-ray fluxes are only possible under sustained high $pp$ collision rates. The VHE decay
however can be then explained by $\gamma\gamma$ absorption. The high $B_{\rm c}$ case is shown in Fig.~\ref{sede} (middle), for which
we have adopted $B_{\rm c}=30$~G. The $\gamma\gamma-e^\pm$ synchrotron contribution is also included in the figure. The SSC and
relativistic Bremsstrahlung levels, not shown, are similar to those of the $pp-e^\pm$ pairs. The multiwavelength SED for protons with
$Q_p(E)\propto E^{-1.5}$ and $B_{\rm c}=0.03$~G is presented in Fig.~\ref{sede} (bottom), and is similar to that of the $Q_p(E)\propto
E^{-2}$ case (the same applies for the case with $B_{\rm c}=30$~G, not shown). 

\begin{figure}
\includegraphics[width=70mm,angle=-90]{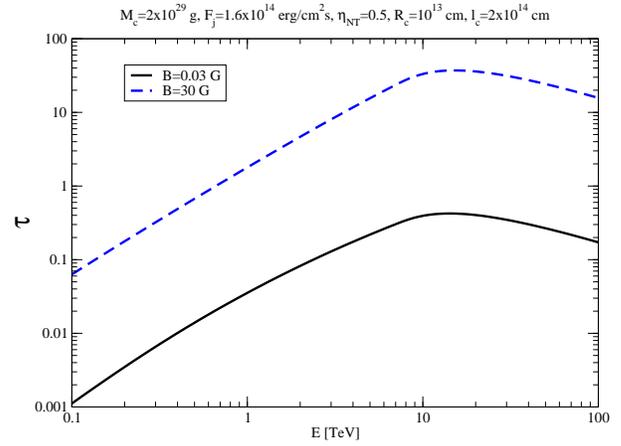}
\caption{The computed $\gamma\gamma$ opacities for the same parameter values of Fig.~\ref{sedpp}, 
and $B_{\rm c}=0.03$ (solid line) and 30~G (dashed line).}
\label{tau}
\end{figure}

\begin{figure}
\includegraphics[width=70mm,angle=-90]{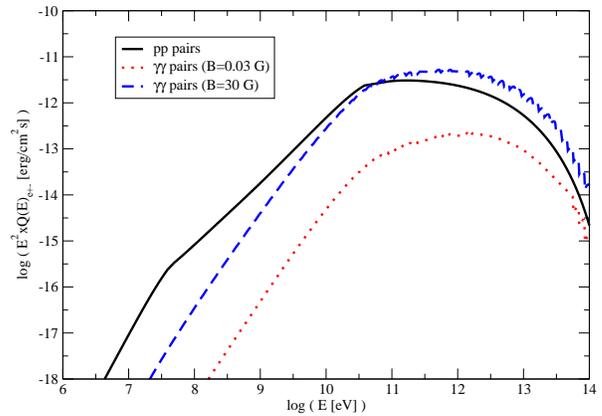}
\caption{The computed SED of the injected $pp-$ and $\gamma\gamma-e^\pm$ pairs for the same parameter values of 
Fig.~\ref{sedpp} ($Q_p(E)\propto E^{-2}$), and $B_{\rm c}=0.03$ (dotted line) and 30~G (dashed line).
{ Units are the same as in Fig.~\ref{sedpp} to facilitate the comparison of the $pp-e^\pm$ and $pp-$gamma-ray SEDs.}}
\label{pairs}
\end{figure}

To illustrate further the impact of $B_{\rm c}$ on gamma-ray suppression and the cloud (secondary) leptonic population, we
present in Figs.~\ref{tau} and \ref{pairs} the opacities to gamma rays, and the SED ($E^2\,\dot{N}(E)$) of the injected $pp-$
and $\gamma\gamma-e^\pm$ pairs, for the two $B_{\rm c}$-values adopted. It is worth noting that $\gamma\gamma$ absorption
softens the VHE spectrum in Fig.~\ref{sede} (middle), in good agreement with observations in the decay epoch
\citep[][]{veritas11}. Note also that the non-absorbed VHE lightcurve peak is broad enough to accommodate the two days
separation between the VHE maximum and the X-ray observation. 

\subsection{Thermal emission from the cloud}

As shown in \cite{bab10}, the shocked cloud is initially optically thick to its own radiation. Given the high cloud pressure, 
$\sim p_{\rm j}$, very high UV photon densities will completely quench all the VHE radiation. With expansion, the cloud
radiation field quickly  dilutes and heats up, with a fast decrease of the $\gamma\gamma$ opacity. At the peak of the flare,
the cloud is fully ionized, moderately optically thin ($\tau_{e\gamma}\approx 0.6\,r_{\rm c,14}\,n_{\rm c,10}$; 
$n_{\rm c,10}=n_{\rm c}/10^{10}\,{\rm cm}^{-3}$),
and
emitting through free-free radiation with temperature $\sim 10^{10}$~K and luminosities of $\sim 10^{41}$~erg~s$^{-1}$. The
optical depth for TeV photons can be estimated as \citep[see][for further details]{bab10}:
\begin{equation}
 \tau_{\gamma\gamma,\rm therm}\approx 10^{-3} r_{\rm c,14}^2\,n_{\rm c,10}^2 T_{10}^{-1/2}.
\end{equation}
The opacity due to thermal radiation can be therefore neglected. The thermal radiation becomes important at MeV
energies around the VHE maximum, when the jet-cloud energy transfer is most efficient. 

\section{Discussion}\label{disc}

The interaction of a gas cloud, or the atmosphere of a disrupted RG, with the base of an AGN jet leads to the formation of an
interaction region. There, jet energy can be dissipated in the form of relativistic protons, which can penetrate into the
cloud, initially dense enough to render $pp$ interactions efficient. This yields gamma rays and other secondary particles, in
particular $e^\pm$ pairs. The energy transfer into the cloud, through MHD interactions (and injection of cosmic rays), leads
to a quick expansion that increases the covered section of the jet and thereby the gamma-ray emission. Around the point when
$pp$ collisions become optically-thin, the gamma rays reach their maximum after a quite sharp rise. After that moment, $pp$
collisions become strongly inefficient quenching the gamma-ray emission. The drop of the gamma-ray flux can be even more
abrupt because of the cloud leaving the jet, and/or $\gamma\gamma$ absorption. Adopting $M_{\rm c}\sim 10^{29}$~g and $L_{\rm
j}\approx 5\times 10^{44}$~erg~s$^{-1}$, the model can reproduce rather nicely the lightcurve and the spectrum of the VHE
flare detected in M87 in April 2010.

At the VHE peak, a low $B_{\rm c}$ is required to avoid gamma-ray absorption in the secondary synchrotron field. This is not
problematic, since the cloud may be initially weakly magnetized. The enhanced X-ray flux few days after the VHE maximum is consistent
with our model if $B_{\rm c}$ increases, e.g. through cloud-jet mixing or complex MHD processes. A clear prediction of the proposed
model is that, at the highest VHE fluxes of an event like the April 2010 one, to avoid absorption, X-rays should not be significantly
enhanced regardless their origin, either primary or secondary leptons. 

The model presented here cannot be applied for particles whose energy evolution timescale is either longer, or very sensitive to the
late dynamical evolution of the cloud, which prevents to derive predictions in radio. However, the compactness of the source already
shows that the radio emission will be likely self-absorbed during the flare. In the IR/optical, the computed fluxes are below the
observed ones even for a high $B_{\rm c}$-field. Later, primary and secondary leptons may non-negligibly contribute to the radio and
IR/optical emission, but a proper account of these radiation components would require a very detailed model of the dynamical and
radiative properties of the interacting flows. In any case, in what concerns the April 2010 flare, the radio and optical emission did
not seem to correlate with the VHE emission \citep{hess11}.

The hard proton spectrum adopted implies rather modest GeV fluxes, although if the injection proton spectrum $\propto E^{-2}$
went down to $\sim 1$~GeV in proton energy, the $0.1-10$~GeV luminosity would be a factor of a few times higher than the
year-averaged flux found by {\it Fermi}. However, for these fluxes and the sensitivity of this instrument, it seems difficult
to probe day-scale variability. 

It is important to emphasize that, because of the lack of beaming and the limited jet luminosities in local Universe AGN,
single interactions of clumps with jets in misaligned jet sources \citep[for blazars, see][]{babkk10} are hard to detect.
Only very nearby objects, like M87, Cen~A, and probably NGC1275, could provide detectable radiation of this origin. More
distant though still local objects may be also detectable, since more clumps are available at further distances from the
SMBH. This emission would appear as persistent though \citep[e.g.][]{abr10,bpb12}, and the modest jet power of the potential
candidates would probably require rather long observations. 

\section*{Acknowledgments}
We thank an anonymous referee for his/her constructive comments and suggestions.
We would like to thank Frank Rieger for useful discussions, and
Martin Raue for supplying the VHE data.
The research leading to these results has received funding from the European
Union Seventh Framework Program (FP7/2007-2013) under grant agreement
PIEF-GA-2009-252463. V.B.-R. acknowledges support by the Spanish 
Ministerio de Ciencia e Innovaci\'on (MICINN) 
under grants AYA2010-21782-C03-01 and FPA2010-22056-C06-02.  


\end{document}